# The Future of the European VLBI Network

**Huib Jan van Langevelde**[1]

*The Joint Institute for VLBI in Europe*
*Postbus 2, 7990 AA Dwingeloo, the Netherlands*
*And Sterrewacht Leiden, Leiden University*
*Postbus 9513, 2300 RA Leiden, the Netherlands*
*E-mail:* `langevelde@jive.nl`

Guided by the recently published science case for the future of European VLBI, EVN2015, a roadmap for the future of the EVN is sketched out in this paper. The various desired technical improvements are being discussed with an emphasis on the role of e-VLBI. With this innovation new scientific capabilities are introduced. In this way the EVN is also positioned as an interesting platform for exercising new techniques and operational models, complementary to other SKA pathfinders. In return, the technology development for the SKA can have a positive impact on the scientific capabilities of VLBI, for example on the development of a next generation correlator, capable to process much larger data-rates. The development of cheap, frequency agile antennas can also be of great importance for VLBI. This adds to the potential for maintaining a Northern hemisphere, global VLBI array in the SKA era.

[1] Speaker

# 1. Introduction

The EVN is developing a long-term roadmap for the next decade. There is an obvious need for such a long-term vision, as funding agencies are actively defining scientific priorities on a European scale, for example in the ASTRONET roadmap exercise (http://www.astronet-eu.org/IMG/pdf/Astronet-Book.pdf). The VLBI community must set its priorities in view of the global ambition of radio astronomers all over the world to develop the Square Kilometer Array (SKA).

Recently, the EVN has developed and published a science vision document called EVN2015 (http://www.evlbi.org/publications/publications.htm). This document sets the scientific priorities for the evolution of the EVN. Based on these scientific objectives, the EVN board of directors has taken the initiative to develop a roadmap in which the technical developments are to be outlined. This exercise also requires a discussion of funding options, as well as matters of operations, science policy and governance of the EVN and JIVE.

While this exercise is still in a very early stage, I will discuss these matters from a more or less personal perspective in this paper. The topics of this presentation range from digital components to antennas; they are being discussed here without having been prioritized yet.

# 2. Scientific motivation

The EVN has developed a particularly strong science case, which is addressing a breadth of astronomical topics. I will discuss a number of specific topics. In the classical area of VLBI research, jets in Active Galactic Nuclei, important progress can be anticipated with a boost in continuum sensitivity, which will also bring radio-quiet AGN within range. This is expected to yield insight in the fuelling and life cycle of AGN. The same boost in sensitivity will also allow the EVN to play a vital role in studying the AGN/starburst connection in cosmological fields. In these areas the VLBI sensitivity must be improved to maintain its complementary role with the EVLA and e-MERLIN facilities by providing high-resolution diagnostics at a similar sensitivity. More sensitivity, especially at higher frequency, will allow VLBI to probe jet physics close to the black hole event horizon. This is particularly relevant in combination with Space VLBI [4][5].

Increases in bandwidth alone will not be sufficient to improve the statistics on extragalactic absorption studies and mega-masers. For this important branch of research, sensitivity improvements are expected from expanding the number of antennas that participate in VLBI. Indeed various new antennas are planned in Europe and beyond (see below). Additionally, a dramatic increase in the number of antennas should be an ambition, as it is crucial for improving the imaging capabilities and obtaining high fidelity pictures from VLBI. Improvements in the receiver agility and robustness, for example by multiple bit recording, are also important for this topic, and indeed for all spectral line projects.

In Galactic astronomy, the study of transient radio sources has already taken advantage of the development of real-time VLBI, but currently the samples under study are limited in size, until much better sensitivity becomes available. Together with large-scale survey instruments, like LOFAR [3], providing many more triggers, this can be expected to be a booming industry in a matter of years. Pulsar capabilities must feature prominently as VLBI pulsar astrometry will be in demand for population studies that harvest the results from a number of SKA pathfinders.

In maser research, the EVN is already providing unique insight in high-mass star formation, through methanol observations [1][13]. In the coming years the EVN will have more capabilities at higher frequencies, providing access to more maser transitions, important for both star formation and studies of evolved stars. In particular, VLBI is playing an important role in determining fundamental distances[6][11], which will be helped by larger bandwidth to reach nearby, but typically faint extragalactic calibrators. Flexibility in the recording and correlation setup is required for this.

Finally, it has been demonstrated that VLBI offers very interesting scientific applications for planetary Space missions, by providing accurate positions of spacecraft. Here too, real time processing is an important asset for the VLBI system of the future.

These science cases call for a development plan that increases the sensitivity and flexibility of the EVN. Besides an obvious requirement to push the bandwidth of the EVN, there is also clearly an interest to move operations to higher frequencies, like 22 and 43 GHz, where there is also more available bandwidth. The science case calls for new VLBI elements in the next decade to provide sensitivity and imaging capabilities at these frequencies. A next generation correlator must provide capacity for processing these higher bandwidths, but also more flexibility for pulsar gating, many bit processing and astrometry.

## 2.1 e-VLBI

In the last few years e-VLBI has evolved from an experimental technique, connecting a small number of telescopes at modest bandwidth into an operational astronomical service with competitive sensitivity and imaging capabilities. The project has been supported by the EC through the FP6 Integrating activity EXPReS[2], which has stimulated a large-scale collaboration between EVN technical staff and European Research Network providers.

The initial argument for developing e-VLBI has been the desire to use VLBI on transient phenomena, being able to access the data on variable sources on the timescale that they vary. Indeed, opening the parameter space accessible to VLBI has turned up some very interesting results, for example [7][8][10]. However, the project has also demonstrated that e-VLBI is not only a technique to do rapid response science. Because of its real-time nature it is also much more robust against failure, which will be noticed immediately and can be addressed instantaneously (Figure 1). Since dedicated light-paths are in use, this is true, even though the connectivity adds potentially an extra layer of complexity [12]. The various demonstrations have also taken away the scepticism whether e-VLBI could ever work for intercontinental baselines, showing fringes between telescopes on different continents simultaneously.

For future VLBI the connectivity promises to be able to deliver larger and larger bandwidth with the 10Gbps standard already operational in many places. Besides reducing the delivery time to the astronomer, e-VLBI can also help in enhancing the flexibility of the networks by avoiding the complex logistics that are involved in shipping scarce recording media. Currently e-VLBI is realised in a large-scale collaboration with the connectivity providers and therefore it is hard to evaluate the economics of connected VLBI in comparison to the transport of magnetic media. Clearly in the long run it is expected that real-time connectivity will be cheaper and more environmentally friendly.

---

[2] EXPReS is an Integrated Infrastructure Initiative (I3), funded under the European Commission's Sixth Framework Programme (FP6), contract number 026642.

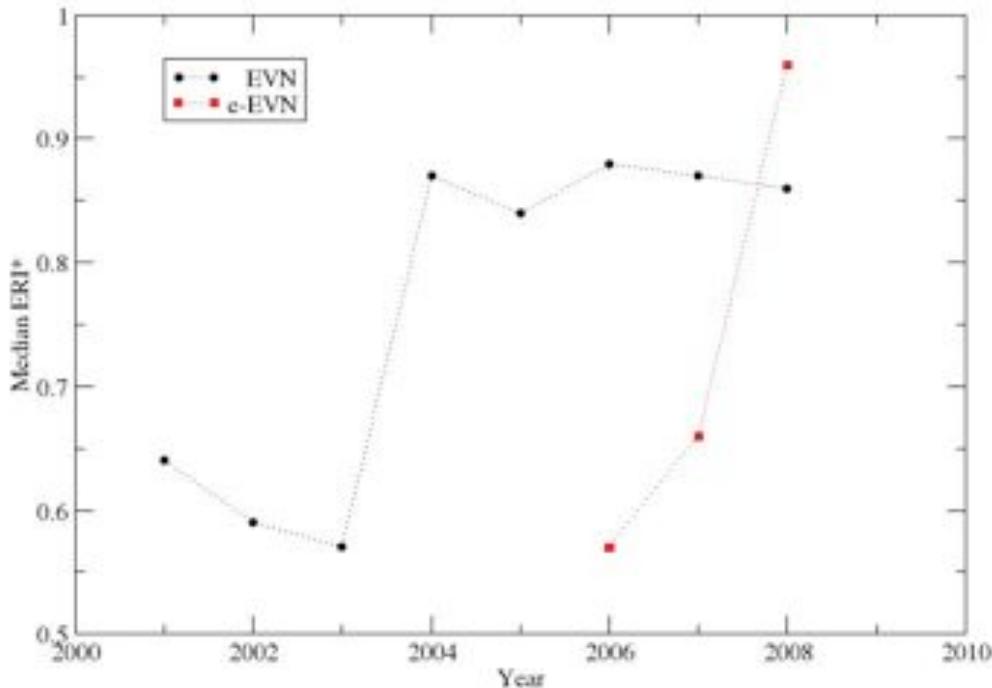

*Figure 1. The EVN Reliability index is a measure of the fraction of data taken succesfully from a scheduled experiment. The averaged ERI for normal, receorded EVN data (black) can be seen to improve dramatically with the introduction of disk recording (2004). The ERI for e-VLBI, albeit based on a small number of relatively simple experiments, is quicly reaching a very promising level.*

One way in which e-VLBI is less flexible than recorded VLBI is that its current operations do not allow re-correlations, which are for example needed to accommodate large numbers of telescopes or spectral line experiments that need mixed bandwidths. It is noted that in principle all of these are limitations of the current correlator, but even so it will be useful to introduce buffering of the VLBI data in various stages in order to combine the best aspects of recorded and real-time operations.

## 3. Bandwidth increase

The EVN has successfully pushed its sensitivity to 1024 Mbps in the past few years. The Mk4 recording system, originally tape-based, was already designed to digitize $16 \times 16$MHz bands. Since the introduction of Mk5 disk-based recording, it has been possible to offer this capability for priority science observations, depending on the availability of disk capacity. Recently this data-rate has also been reached for e-VLBI observations [12]. In 2010 it should be possible to push the sensitivity to 4 Gbps. There are various options for recording such data-rate capacities using modern computer components [14]. More critical seems to be the introduction of the new filters and digitizers, the DBBC system in the case of the EVN. This system is needed for any new telescope that joins the array and eventually for all stations to go beyond 1024Mbps. Moreover, the EVN bandwidth is limited by the inherent width of the IF systems of the telescopes, and this must be addressed at many stations in the future.

Eventually, there is the ambition to go to even higher bandwidths, which will be mostly useful at higher frequencies. Here possibly up to 2 GHz bandwidth could be realized in each polarization, calling for at least 16 Gbps data-rates. Higher data-rates (128 Gbps) can be utilized by having up to 8-bit signal representation, which can overcome some of the adverse effects introduced by radio-frequency interference, notably at L-band.

## 4. Next Generation correlator

The current EVN Mk4 correlator at JIVE that is at the heart of the EVN operations is capable of processing 16 stations at 1024 Mbps. Even though its capacities are still being enhanced [2], it is already necessary to invoke multiple pass correlation for a considerable fraction of experiments. This is required for large global experiments that use more than 16 stations at one time, irrespective of the recording rate. Multiple-pass correlation is also requested by users who want extremely high spectral resolution, or spectral processing of more than one sub-band. Notably this occurs when continuum sensitivity is required at the same time, for example in phase referencing.

Higher than 1024 Mbps data rates can in principle be supported by multiple pass correlation, provided all sample rates are below 32 Msamples per second. However, it is obviously not possible to accommodate high data-rates for e-VLBI in this way, unless the data is buffered at the correlator. Similarly, e-VLBI is of limited use for a considerable fraction of spectral line projects.

For the future a large-scale new EVN data processor is therefore required. Looking at the above specifications and the EVN (and global) ambitions to have more stations (below), one can anticipate the need to process 32 stations (a factor 4 compared to the current), 16 Gsamples per second (a factor 16) and maybe a 4 to 8 times better spectral resolution than the current EVN data processor. Overall the aim for 2015 must be to develop a correlator that is at least a hundred times more powerful than the current data-processor. This calls for a machine of the same order of size as the correlators for other SKA precursors (MeerKAT, ASKAP) and pathfinders (e-MERLIN, EVLA, APERTIF).

Various VLBI arrays around the world are adopting software correlators that can perform VLBI processing on commodity computer clusters. This approach is superior in flexibility and accuracy over hardware implementations and on moderately sized clusters these perform similarly as the current hardware-based machines. At JIVE the evaluation of software correlators is done in the context of the SCARIe/FABRIC programme and it is recognized that this option is interesting for the near future needs of VLBI. Taking future energy constraints into consideration, we are actively pursuing for correlation on FPGA based architectures, reaching a similar conclusion as other SKA pathfinder projects.

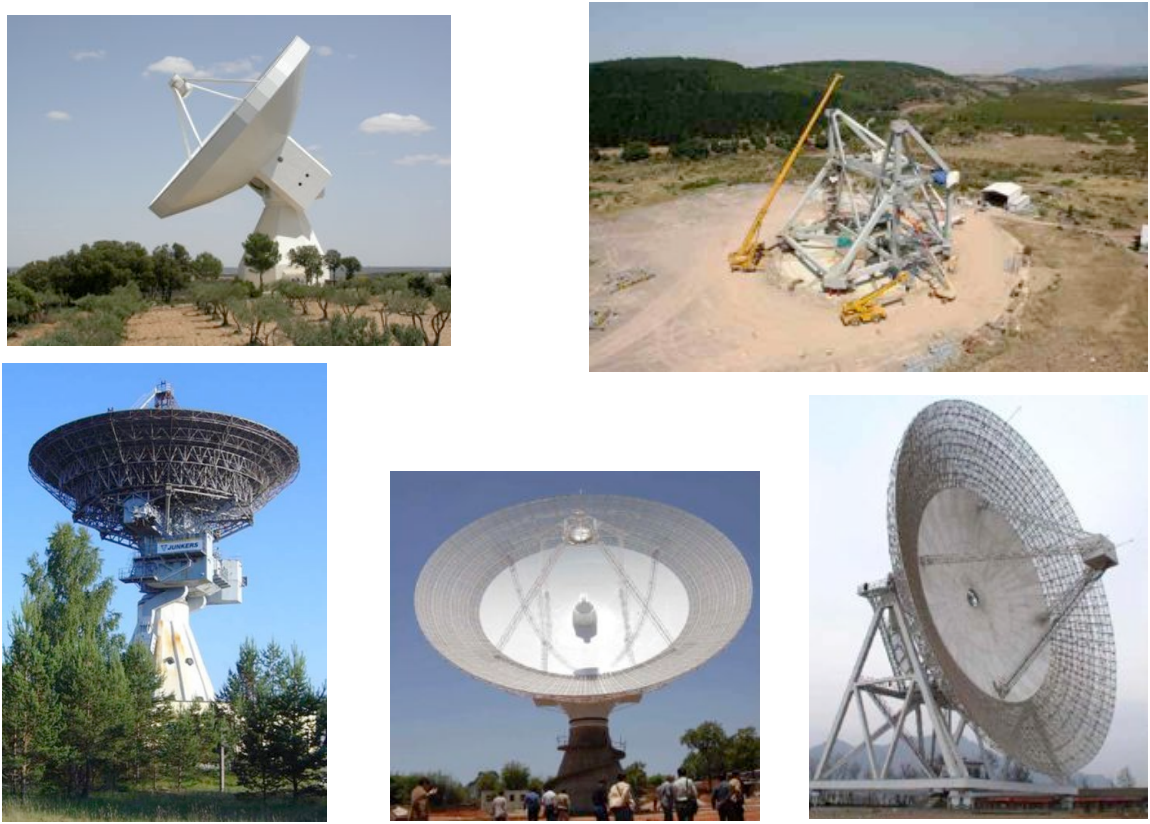

*Figure 2. From the left top corner, the new Yebes 40m, the site of the Sardina telescope (summer 2008), the Ventspils (Latvia) telescope, and Kunming and Miyun in China*

## 5. Telescopes

One of the main motivations to increase the correlator capacity is the increasing number of telescopes. In global observations the number of telescopes can already go beyond 16. In the near future it will be possible to observe with EVN+MERLIN, using all stations. Moreover, recently the new Yebes 40m (Spain) telescope has started observing with the EVN. The first tests with Ventspils 32m (Latvia) are being carried out. In Italy the Sardinia Radio Telescope (64m) is expected to have first light in 2009[9]. In China the new telescopes of Miyun (50m) and Kunming (40m) have participated in the Chang'e experiments (Figure 2). These new, big dishes will contribute to the sensitivity; in particularly for spectral line the importance of pure collecting area is enormous. Moreover, most of the new antennas are extending the high frequency capability of the EVN in the 22 and 43 GHz range. But not only sensitivity but certainly also image fidelity is improving with a larger number of antennas.

Beyond these telescopes under construction, the EVN is also actively working with telescopes in Russia and the Ukraine to improve the science capabilities. In a number of other countries there are ambitions to construct telescopes that can be of great value to the *uv*-coverage of our array.

Ultimately, it is important to transfer VLBI into an imaging instrument with great fidelity, in order to complement the high-resolution pictures that can be expected to emerge in the next decades at other wavelengths. This can be achieved by having VLBI arrays with an order of magnitude more telescopes. The developments in the SKA context of cheap, relatively small antennas, may offer interesting options for this. One can imagine the current VLBI arrays augmented with many telescope sites, each equipped with a few (≈6), small (≈12m) antennas (Figure 3). Such antenna clusters can be deployed flexibly, for example with one or two tracking calibrators and the others full time on the target. Such an arrangement would also be the ideal VLBI element for some of the Space science applications that are being considered.

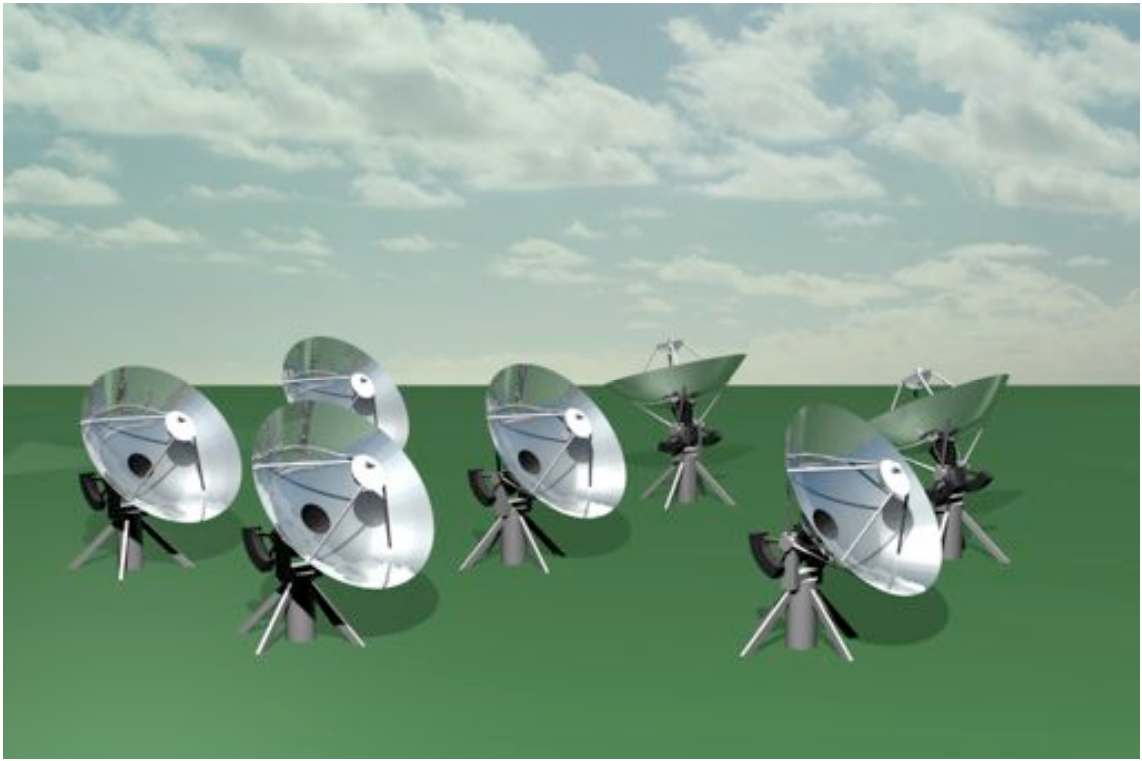

*Figure 3 Artist impression of a possible station configuration that is using modest size telescopes to create a flexible VLBI station (image by P. Boven, JIVE)*

The future also dictates that the existing VLBI arrays become much more flexible in assigning their elements to joint observations. Operational models must evolve in such a way that users can easily define the optimal array in extent, frequency and sensitivity for their observations. It should be possible to flexibly schedule and adapt observations in order to reach a guaranteed level of quality, taking atmospheric conditions into account.

In the EVN 2015 document, a strong science case is presented that addresses fundamental questions in astronomy. This calls for long baseline capabilities that are currently not completely covered in the SKA roadmap, certainly not in its first two phases. This array could be in the Northern hemisphere and encompass the current EVN telescopes. After all, I think there are also other motivations in training and outreach to keep our fantastic telescopes going in the same time that we are designing and building the SKA.